\documentclass[journal=jctcce,manuscript=article, chaptertitle=true]{achemso}
\usepackage{graphicx}
\usepackage{amsmath}
\usepackage{amsfonts}
\usepackage{amssymb}
\usepackage{float}

\author{Tomislav Begu\v{s}i\'{c}}
\altaffiliation{These authors contributed equally.}
\affiliation[EPFL]{Laboratory of Theoretical Physical Chemistry, Institut des Sciences et Ing\'enierie Chimiques, Ecole Polytechnique F\'ed\'erale de Lausanne (EPFL), CH-1015, Lausanne, Switzerland}
\author{Enrico Tapavicza}
\email{enrico.tapavicza@csulb.edu}
\altaffiliation{These authors contributed equally.}
\affiliation[CSULB]{Department of Chemistry and Biochemistry, California State University, Long Beach, 1250 Bellflower Boulevard, Long Beach, California 90840-9507, United States}
\author{Ji\v{r}\'i Van\'i\v{c}ek}
\email{jiri.vanicek@epfl.ch}
\affiliation[EPFL]{Laboratory of Theoretical Physical Chemistry, Institut des Sciences et Ing\'enierie Chimiques, Ecole Polytechnique F\'ed\'erale de Lausanne (EPFL), CH-1015, Lausanne, Switzerland}

\title{Applicability of the thawed Gaussian wavepacket dynamics to the calculation of vibronic spectra of molecules with double-well potential energy surfaces}

\graphicspath{{./figures/}}

\begin{document}

\begin{tocentry}
\includegraphics[scale=0.5]{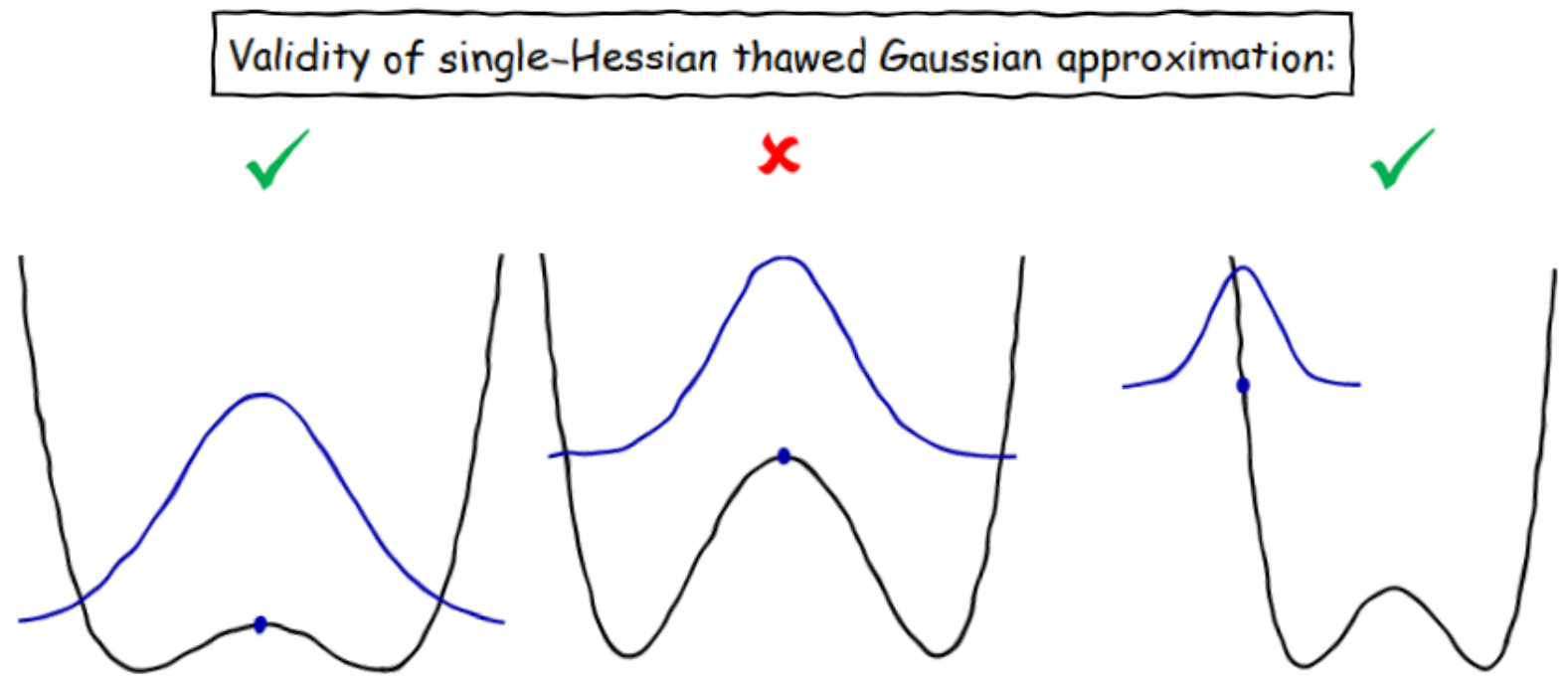}
\end{tocentry}

\begin{abstract}
Simulating vibrationally resolved electronic spectra of anharmonic systems, especially those involving double-well potential energy surfaces, often requires expensive quantum dynamics methods. Here, we explore the applicability and limitations of the recently proposed single-Hessian thawed Gaussian approximation for the simulation of spectra of systems with double-well potentials, including 1,2,4,5-tetrafluorobenzene, ammonia, phosphine, and arsine. This semiclassical wavepacket approach is shown to be more robust and to provide more accurate spectra than the conventional harmonic approximation. Specifically, we identify two cases in which the Gaussian wavepacket method is especially useful due to the breakdown of the harmonic approximation: (i) when the nuclear wavepacket is initially at the top of the potential barrier but delocalized over both wells, e.g., along a low-frequency mode, and (ii) when the wavepacket has enough energy to classically go over the low potential energy barrier connecting the two wells. The method is efficient and requires only a single classical ab initio molecular dynamics trajectory, in addition to the data required to compute the harmonic spectra. We also present an improved algorithm for computing the wavepacket autocorrelation function, which guarantees that the evaluated correlation function is continuous for arbitrary size of the time step.
\end{abstract}

\section{Introduction}

Electronic and photoelectron spectroscopies are regularly used for studying electronic states of molecules and the associated potential energy surfaces.\cite{Barone_Puzzarini:2021,michaelahollas1993determination} To interpret or simulate vibronic features of measured spectra, researchers often invoke the harmonic approximation,\cite{Santoro_Barone:2007,Santoro_Barone:2008,Barone_Santoro:2009,Niu_Shuai:2010,AvilaFerrer_Santoro:2012,Baiardi_Barone:2013,Benkyi_Sundholm:2019,de2021first} which is sometimes valid and greatly simplifies the calculations. However, there are many systems for which this approximation fails, including floppy molecules that exhibit large-amplitude motion along bending or torsional degrees of freedom.\cite{michaelahollas1993determination,Borrelli_Peluso:2006, Borrelli_Peluso:2008,Peluso_Capobianco:2009,Capobianco_Peluso:2012,Wehrle_Vanicek:2015,Baiardi_Barone:2017} The anharmonicity of the potential energy is prominent in these modes due to a large difference between the equilibrium geometries of the initial and final electronic states.

Many exact and approximate methods exist to model vibrationally resolved electronic and photoelectron spectra of anharmonic systems.\cite{Barone_Puzzarini:2021,Prlj_Curchod:2021,Zuehlsdorff_Isborn:2019,Srsen_Heger:2020,Zuehlsdorff_Isborn:2021,Shedge_Isborn:2021,Srsen_Slavicek:2021} Time-independent approaches\cite{Mok_Dyke:2000,Luis_Kirtman:2004,Luis_Christiansen:2006,Bowman_Meyer:2008,Koziol_Krylov:2009,Yang_Lin:2012,Meier_Rauhut:2015,Egidi_Barone:2017} are based on solving the time-independent Schr\"{o}dinger equation for the vibrational degrees of freedom and on computing the Franck--Condon factors, which define the intensities of spectral peaks. For larger molecules, however, the number of non-negligible Franck--Condon factors and associated vibrational eigenstates is intractable. The large couplings between the modes often require a multi-dimensional anharmonic treatment,\cite{Luis_Kirtman:2004,Luis_Christiansen:2006} which further increases the computational cost of time-independent approaches. In addition, the computational effort of evaluating individual vibronic transitions is wasted if these are unresolved due to spectral broadening effects. This inefficiency is avoided in an alternative, time-dependent approach,\cite{Heller:1981a,book_Mukamel:1999,book_Tannor:2007} in which the spectrum is computed as a Fourier transform of an appropriate time correlation function, because low-resolution spectra require only short-time quantum-dynamical calculations. Moreover, the time-dependent approach provides an excellent starting point for efficient and accurate approximations.

Most quantum dynamics methods require the full potential energy surface,\cite{Kosloff_Kosloff:1983,Meyer_Cederbaum:1990,Chen_Guo:1999,Burghardt_Giri:2008,book_MCTDH,Choi_Vanicek:2019,Roulet_Vanicek:2019} whose evaluation is challenging even for moderately-sized molecules, due to the exponential scaling with the number of degrees of freedom.\cite{Shalashilin_Child:2004} To avoid this difficult task, a number of on-the-fly trajectory-based approaches\cite{Ben-Nun_Martinez:2000,Child_Shalashilin:2003,Tatchen_Pollak:2009,Ianconescu_Pollak:2013,Ceotto_Atahan:2009,Ceotto_Atahan:2009a,Wong_Roy:2011,Saita_Shalashilin:2012,Makhov_Shalashilin:2017,Richings_Lasorne:2015,Gabas_Ceotto:2017,Ceotto_Conte:2017,Bonfanti_Pollak:2018,Werther_Grossmann:2020,Werther_Grossmann:2021} have been developed, where the potential energy can be computed locally and only when needed, using ab initio electronic structure methods. These methods account intrinsically for anharmonic effects, either exactly or approximately. However, many on-the-fly ab initio methods based on multiple trajectories are impractical for use in routine spectra calculations.\cite{Makhov_Shalashilin:2017}

Here, we discuss a practical, single-trajectory approach called the single-Hessian thawed Gaussian approximation,\cite{Begusic_Vanicek:2019} which was developed recently as an efficient way to include anharmonic effects in the calculation of vibrationally resolved electronic and photoelectron spectra. This on-the-fly ab initio method, which is an efficient but more approximate variant of Heller's thawed Gaussian approximation,\cite{Heller:1975,Grossmann:2006,Wehrle_Vanicek:2014} was constructed primarily to treat weak-to-moderate anharmonic effects. Here, we further explore its advantages over the harmonic approximation but also its limitations by investigating systems with double-well potential energy surfaces. We first review the theory of harmonic and thawed Gaussian approximations in Section~\ref{sec:theory}, where we also propose an improved, robust algorithm for computing the autocorrelation function of a Gaussian wavepacket. Then, we study how the harmonic and single-Hessian thawed Gaussian approximations perform in simulating the absorption spectrum of 1,2,4,5--tetrafluorobenzene (TFB), emission and photoelectron spectra of ammonia, and photoelectron spectra of phosphine and arsine. Due to their symmetry, all these systems exhibit double-well potential energy surfaces.

\section{\label{sec:theory}Theory}

\subsection{\label{subsec:spec_theory}Time-dependent approach to vibrationally resolved absorption, emission, and photoelectron spectroscopies}

The absorption cross-section measured in electronic absorption or photoelectron spectroscopies can be computed approximately as\cite{Heller:1981a,book_Mukamel:1999,book_Tannor:2007}
\begin{equation}
\sigma_{\text{abs}}(\omega) = \frac{4 \pi \omega}{3 \hbar c} |\mu_{fi}|^2\text{Re}\int_0^{\infty} dt\, e^{i (\omega +  \omega_{i,g}) t} C(t), \label{eq:sigma_abs}
\end{equation}
where
\begin{equation}
C(t) = \langle \psi_0 | \psi_t \rangle \label{eq:C_t}
\end{equation}
is the autocorrelation function of the nuclear wavepacket
\begin{equation}
| \psi_t \rangle = e^{-i\hat{H}_f t / \hbar} |i,g \rangle,
\end{equation}
$|i,g\rangle$ is the ground vibrational state ($g$) of the initial electronic state ($i$), $\hbar \omega_{i,g}$ is its energy, $\hat{H}_f$ is the vibrational Hamiltonian of the final electronic state, and $\mu_{fi}$ is the electronic transition dipole moment. To arrive at eq~\ref{eq:sigma_abs}, we made the zero-temperature approximation, which assumes that only the ground vibrational state of the initial electronic state is populated, and the Condon approximation, in which the transition dipole moment is assumed to be independent of nuclear coordinates in the region of the initial state. For expressions beyond these approximations, see, for example, refs~\citenum{book_Mukamel:1999,Niu_Shuai:2010,Baiardi_Barone:2013}. Whereas the Condon approximation breaks down for symmetry forbidden electronic transitions (such as in benzene), the zero-temperature approximation cannot be used in large floppy molecules, where low-frequency vibrational modes might already be excited at room temperature. In the semiclassical method that we will use here, one can include finite-temperature effects with the thermofield dynamics\cite{Begusic_Vanicek:2020,Begusic_Vanicek:2021a} and non-Condon effects with the extended thawed Gaussian approximation\cite{Patoz_Vanicek:2018,Begusic_Vanicek:2018a,Vanicek_Begusic:2021}, but for simplicity we will assume both Condon and zero-temperature approximations in the main text and analyze their validity only in the Supporting Information (Section 6). As our main interest is gas phase spectroscopy, we do not consider coupling of the molecule to a bath here.

In electronic absorption spectroscopy, the initial state $i$ is the ground electronic state of the molecule, and the final state $f$ is one of the excited electronic states. In photoelectron spectroscopy, the initial state corresponds to the ground state of the neutral molecule, while the final state is the ground electronic state of the cation. The fluorescence spectrum is defined as the rate of emission per unit frequency:\cite{Niu_Shuai:2010,book_Tannor:2007}
\begin{equation}
\sigma_{\text{em}}(\omega) = \frac{4 \omega^3}{3 \pi \hbar c^3}|\mu_{fi}|^2 \text{Re}\int_0^{\infty} dt e^{i (\omega - \omega_{i,g}) t} C(t)^{\ast},
\end{equation}
where the initial state is the excited electronic state from which the emission occurs, while the final state is the ground electronic state of the molecule.

In all types of spectroscopy discussed in this Section, wavepacket dynamics, appearing in the correlation function of eq~\ref{eq:C_t}, is central and bears the main computational cost. For typical molecules containing tens or hundreds of degrees of freedom, simulating wavepacket dynamics numerically exactly is a challenging, if not an impossible, task. Fortunately, efficient schemes exist to evaluate the correlation function (eq~\ref{eq:C_t}) under reasonable approximations. In the following, we discuss two such approaches.

\subsection{\label{subsec:harmonic}Harmonic approximation}

Within the harmonic approximation, the potential energy is assumed to be a quadratic function of nuclear coordinates obtained by expanding the true potential to the second order about a reference geometry $\xi_{\text{ref}}$:
\begin{equation}
V(\xi) \approx V_{\text{HA}}(\xi) = V(\xi_{\text{ref}}) + V^{\prime}(\xi_{\text{ref}})^{T} \cdot (\xi - \xi_{\text{ref}}) + \frac{1}{2}(\xi - \xi_{\text{ref}})^T \cdot V^{\prime \prime}(\xi_{\text{ref}}) \cdot (\xi - \xi_{\text{ref}}).
\end{equation}
In principle, any molecular geometry can be taken as the reference. However, the most common choice is to expand both initial- and final-state potential energies about their corresponding minima ($\xi_{\text{ref}} = \xi_{\text{eq}}$), which defines the so-called adiabatic harmonic model.\cite{AvilaFerrer_Santoro:2012} The harmonic potential energy can then be rewritten in mass-scaled normal mode coordinates as
\begin{equation}
V_{\text{HA}}(q) = V_{\text{eq}} + \frac{1}{2} q^T \cdot \Omega^2 \cdot q,
\end{equation}
where $V_{\text{eq}} = V(\xi_{\text{eq}})$, $\Omega = \text{diag}(\omega_1, \dots , \omega_D)$ is a diagonal matrix of harmonic frequencies, and $D$ is the number of degrees of freedom. The relation between normal mode coordinates of the initial (labeled $q_i$) and final ($q_f$) electronic states is typically described by the Duschinsky transformation\cite{Duschinsky:1937}
\begin{equation}
q_f = J \cdot q_i + d,
\end{equation}
where $J$ is called the Duschinsky matrix and $d$ the displacement vector. This transformation is an approximation that is valid as long as the rotational and vibrational degrees of freedom can be well separated. Within the adiabatic harmonic approximation and assuming the validity of the Duschinsky transformation, the correlation function can be evaluated analytically according to\cite{Etinski_Marian:2011,Baiardi_Barone:2013,Tapavicza_Sundholm:2016}
\begin{align}
C_{\text{HA}}(t) &= e^{-i V_{f, \text{eq}} t / \hbar} G(t) \label{eq:C_HA}\\
G(t) &= \left[ \frac{\det (2 \Omega_f \cdot S_f^{-1} \cdot \Omega_i)}{\det(L \cdot M)}\right]^{1/2} \exp \left( v_f^T \cdot M^{-1} \cdot v_f - d \cdot B_f \cdot d \right), \label{eq:G_t}\\
S_f &= i \sin(\Omega_{f} t),\qquad
B_f = i \Omega_f \cdot \tan(\Omega_f t / 2), \qquad v_f = J^T \cdot B_f \cdot d  \label{eq:G_t_defs1}\\
M &= J^T \cdot B_f \cdot J + \Omega_i, \qquad
L = J^T \cdot \Omega_f^2 \cdot B_f^{-1} \cdot J + \Omega_i. \label{eq:G_t_defs2}
\end{align}

Let us mention two other ways of constructing the harmonic model for the final-state potential energy surface.\cite{AvilaFerrer_Santoro:2012,Ferrer_Santoro:2013} The first is the vertical harmonic approach, which expands the final-state potential energy about the Franck--Condon geometry, i.e., about the equilibrium geometry of the initial state. The second is the adiabatic shift model, which avoids the evaluation of the final-state Hessian by setting $\Omega_f = \Omega_i$ and $J = I_D$. This model is also known as the displaced harmonic oscillator model because it only accounts for the displacement between the two harmonic potential energy surfaces, but neglects the changes in the mode frequencies and the Duschinsky effects due to inter-mode coupling.

The main computational cost of the harmonic method comes from geometry optimizations and Hessian evaluations in the initial and final electronic states, whereas the cost of evaluating eqs~\ref{eq:C_HA}--\ref{eq:G_t_defs2} is almost negligible. Therefore, the harmonic method is extremely practical and feasible even for rather large molecular systems. However, because it completely neglects anharmonicity effects, this method can be inadequate when the potential energy surfaces are far from harmonic, which is the case with double-well potentials.

\subsection{\label{subsec:tga}Single-Hessian thawed Gaussian approximation}

To improve over the harmonic approximation and include at least some anharmonicity effects, we return to the wavepacket autocorrelation expression (eq~\ref{eq:C_t}). As mentioned in Section~\ref{subsec:spec_theory}, the main computational cost of this approach is related to the time evolution of the nuclear wavepacket, which requires solving the time-dependent Schr\"{o}dinger equation
\begin{equation}
i \hbar |\dot{\psi}_t \rangle = \hat{H} | \psi_t \rangle
\end{equation}
with $|\psi_0\rangle = | i,g \rangle$ and $\hat{H} = \hat{H}_f$. Within the thawed Gaussian approximation,\cite{Heller:1975} the wavepacket is assumed to be a Gaussian
\begin{equation}
\psi_t(q) = \frac{1}{(\pi \hbar)^{D/4} (\det Q_t)^{1/2}} \exp \left\{ \frac{i}{\hbar} \left[ \frac{1}{2} (q - q_t)^T \cdot P_t \cdot Q_t^{-1} \cdot (q-q_t) + p_t^T \cdot (q - q_t) + S_t \right] \right\}, \label{eq:gwp}
\end{equation}
where $q_t$ and $p_t$ are $D$-dimensional expectation values of the position and momentum, $Q_t$ and $P_t$ are $D \times D$ complex matrices satisfying\cite{Hagedorn:1980,Faou_Lubich:2009,Lasser_Lubich:2020}
\begin{align}
Q_t^T \cdot P_t - P_t^T \cdot Q_t &= 0, \label{eq:sympl_cond_1} \\
Q_t^{\dagger} \cdot P_t - P_t^{\dagger} \cdot Q_t &= 2 i I_D, \label{eq:sympl_cond_2}
\end{align}
$S_t$ is a real, time-dependent scalar, and $I_D$ is the $D \times D$ identity matrix. Following Heller,\cite{Heller:1975} the potential energy is replaced by its local harmonic approximation about the center of the wavepacket
\begin{equation}
V(q) \approx V_{\text{LHA}}(q) = V(q_t) + V^{\prime}(q_t)^T \cdot (q - q_t) + \frac{1}{2} (q -q_t)^T \cdot V^{\prime \prime}(q_t) \cdot (q - q_t).
\end{equation}
Then, the Gaussian wavepacket (eq~\ref{eq:gwp}) solves the Schr\"{o}dinger equation
\begin{equation}
i \hbar |\dot{\psi}_t \rangle = \Big( \frac{1}{2}\hat{p}^T \cdot m^{-1} \cdot \hat{p} + \hat{V}_{\text{LHA}} \Big) | \psi_t \rangle
\label{eq:psi_dot}
\end{equation}
if the parameters of the Gaussian satisfy the following system of first-order ordinary differential equations
\begin{align}
    \dot{q}_t = m^{-1} \cdot p_t, &\qquad \dot{p}_t = - V^{\prime}(q_t) \label{eq:qp_dot}\\
    \dot{Q}_t = m^{-1} \cdot P_t, &\qquad \dot{P}_t = - V^{\prime \prime}(q_t) \cdot Q_t, \label{eq:QP_dot}\\
    \dot{S}_t = \frac{1}{2}p_t^T &\cdot m^{-1} \cdot p_t - V(q_t) \label{eq:S_dot}
\end{align}
where $m$ is a real symmetric mass matrix. To define the initial conditions, we choose to work in the mass-scaled normal modes of the initial electronic state, for which we invoke the harmonic approximation. Then, the initial wavepacket $| \psi_0 \rangle = | i,g \rangle$ is a Gaussian parametrized by $q_0=0$, $p_0 = 0$, and
\begin{equation}
Q_0 = (\Omega_i/\hbar)^{-1/2}, P_0 = i (\Omega_i / \hbar)^{1/2},
\end{equation}
while the mass matrix reduces to $m = I_D$.

According to eqs~\ref{eq:qp_dot}--\ref{eq:S_dot}, the center of the wavepacket follows a classical trajectory, the propagation of $Q_t$ and $P_t$ requires the Hessians along this classical trajectory, and $S_t$ is the associated classical action. Because the propagation of the wavepacket parameters requires only local potential energy information evaluated at $q_t$, the method is suitable for an on-the-fly ab initio implementation. However, the cost of evaluating Hessians using ab initio electronic structure can be prohibitive in certain cases, e.g., for larger molecules or for highly accurate but costly electronic structure methods. To avoid these limitations, a number of efficient updating,\cite{Ceotto_Hase:2013,Conte_Ceotto:2019} interpolation,\cite{Wehrle_Vanicek:2014,Wehrle_Vanicek:2015} or machine-learning\cite{Laude_Richardson:2018,Gandolfi_Ceotto:2021} schemes were proposed. To this end, two of us recently proposed a crude but practical approach\cite{Begusic_Vanicek:2019} that replaces the Hessians evaluated along the trajectory by a single Hessian evaluated at a reference geometry, i.e., 
\begin{equation}
V_{\text{LHA}}(q) \approx V_{\text{SHA}}(q) = V(q_t) + V^{\prime}(q_t)^T \cdot (q - q_t) + \frac{1}{2} (q -q_t)^T \cdot V^{\prime \prime}(q_{\text{ref}}) \cdot (q - q_t). \label{eq:V_SHA}
\end{equation}
This single-Hessian method was shown to be almost as accurate as the original thawed Gaussian approximation in several model and realistic systems, and was used to simulate both steady-state\cite{Prlj_Vanicek:2020,Begusic_Vanicek:2021a} and time-resolved spectra.\cite{Golubev_Vanicek:2020,Begusic_Vanicek:2021} As in the harmonic approximation, there are different possible choices for choosing the reference Hessian, including the adiabatic and vertical Hessians of the final state or the adiabatic Hessian of the initial state (labeled ``initial Hessian'').

\subsection{\label{subsec:C_t_half_time}Efficient and robust evaluation of the correlation function}

Let us return to the evaluation of the correlation function (eq~\ref{eq:C_t}). Assuming that the initial wavepacket is real in position representation and that the Hamiltonian is independent of time and the state, the correlation function can be rewritten as\cite{Engel:1992,Manthe_Cederbaum:1992,Beck_Jackle:2000}
\begin{equation}
C(t) = \int dq \psi_{t/2}(q)^2. \label{eq:C_t_half_int}
\end{equation}
Although these assumptions hold for the wavepacket propagated exactly under the vibrational Hamiltonian of the final electronic state, in the (single-Hessian) thawed Gaussian approximation, the approximate potential energy depends on the parameter $q_t$ of the Gaussian wavepacket, and, therefore, correlation functions of eqs~\ref{eq:C_t_half_int} and \ref{eq:C_t} are not exactly equal. Clearly, eq~\ref{eq:C_t_half_int} is two times more efficient because it requires the propagation of the wavepacket only up to time $t/2$. In the Supporting Information, we show that the difference between the correlation functions computed with eqs~\ref{eq:C_t} or \ref{eq:C_t_half_int} is small compared to the overall error of the thawed Gaussian approximation. Below, we discuss another advantage of eq~\ref{eq:C_t_half_int} over eq~\ref{eq:C_t} in the context of Gaussian wavepacket dynamics.

For a Gaussian wavepacket $\psi_t(q)$ (eq~\ref{eq:gwp}), the integral of eq~\ref{eq:C_t_half_int} can be evaluated as
\begin{equation}
\int dq \psi_{t}(q)^2 = \det(-i P_t \cdot Q_t)^{-1/2} e^{-i(p_t^T \cdot Q_{t} \cdot P_{t}^{-1} \cdot p_t - 2S_t)/\hbar}.
\end{equation}
To ensure the continuity of the correlation function, special care must be taken when evaluating the complex square root in the prefactor. One approach is to compare the phase (argument) of the complex determinant to the phase in the previous step and adjust it by $\pm 2 \pi$ if the difference is greater than $\pi$ or smaller than $-\pi$.\cite{Begusic:2021} An equivalent approach is to multiply the prefactor by $-1$ each time the determinant under the square root crosses the branch cut. Both approaches rely on the assumption that the phase of the complex determinant will not change by more than $\pi$ between two steps of the dynamics. Standard step sizes (about 40--50 times smaller than the period of the fastest oscillation in the system) are often small enough to guarantee that the phase changes by less than $\pi$ in one step. However, for high-dimensional systems, the phase of the determinant, which is the sum of arguments of the matrix eigenvalues, can change drastically even in a single step. Then, the standard step size could be insufficiently small to ensure the continuity of the correlation function.

Here, we implemented an alternative method to evaluate the complex square root of the prefactor continuously. The denominator in the prefactor is first rewritten as
\begin{equation}
\det(-i P_t \cdot Q_t)^{1/2} = \det(Q_t) \det(-i P_t \cdot Q_t^{-1})^{1/2}.
\end{equation}
Then, we note that the matrix $\alpha_t = -i P_t \cdot Q_t^{-1}$ is symmetric (follows from eq~\ref{eq:sympl_cond_1}) and that its real part is positive definite (follows from eq~\ref{eq:sympl_cond_2}).\cite{Faou_Lubich:2009,Begusic_Vanicek:2019} Finally, we show in the Supporting Information that
\begin{equation}
\det(\alpha_t)^{1/2} = \prod_{j=1}^{D} \sqrt{\lambda_j}, \label{eq:alpha_t_det_sqrt}
\end{equation}
where $\sqrt{z}$ denotes the principal square root of a complex number $z$ and $\lambda_j$ are the eigenvalues of $\alpha_t$. In other words, by evaluating the complex square root of the determinant of $\alpha_t$ as the product of principal square roots of its eigenvalues, we can evaluate the correlation function without any constraint on the step size. Although the diagonalization of a matrix is computationally more expensive than the evaluation of the determinant, the additional cost is negligible compared to the cost of ab initio calculations. Note that the restriction on the step size cannot be avoided if the correlation function is evaluated according to eq~\ref{eq:C_t} because $Q_t$ appearing in the prefactor $\det(Q_t)^{1/2}$ is a general complex matrix. Therefore, eq~\ref{eq:C_t_half_int} is not only a more efficient but also a more robust way of computing the wavepacket autocorrelation function with the thawed Gaussian wavepacket dynamics.

\section{Computational details}
All electronic structure calculations were carried out with TURBOMOLE v7.3.1,\cite{TURBOMOLE} using the second-order approximate coupled cluster singles and doubles (CC2)\cite{Hattig00,Hattig02,Hattig03} method with either the aug-cc-pVTZ basis set (for NH$_3$, PH$_3$, and AsH$_3$) or the def2-TZVP basis set\cite{Weigend2005} (for TFB). Molecular geometries (reported in the Supporting Information) were optimized using a convergence criterion of 10$^{-4}$ for Cartesian gradients and 10$^{-7}$ for the energy. Hessian and frequency calculations were carried out with the NumForce module,\cite{Deglmann2002b} whereas the ab initio trajectories needed for the thawed Gaussian simulations were computed with the frog module\cite{Elliott2000}. For each spectrum, a single trajectory was run for 1000 steps in the final electronic state starting from the equilibrium geometry of the initial electronic state with zero initial velocity. Namely, for vibrationally resolved photoelectron spectra (NH$_3$, PH$_3$, and AsH$_3$), dynamics was carried out in the cationic state using the equilibrium ground-state geometry of the neutral state as initial coordinates. In the case of the NH$_3$ emission spectrum, the trajectory evolved in the ground electronic state using the equilibrium structure of the first excited singlet state (S$_1$) as initial structure. For the absorption spectrum of TFB, the trajectory was computed in the S$_1$ state starting from the ground-state optimized geometry. A time step of 10 au ($\approx 0.24$~fs) was used in all ab initio molecular dynamics simulations.

Adiabatic harmonic vibronic spectra were computed with the radless module\cite{Tapavicza_Sundholm:2016,Tapavicza:2019} (method described in Section~\ref{subsec:harmonic}). To this end, the time correlation function was propagated for 100000 au using a time step of 1 au. The single-Hessian thawed Gaussian method, described in Sections~\ref{subsec:tga} and \ref{subsec:C_t_half_time} was implemented in a development version of TURBOMOLE and will be available in a future official release of TURBOMOLE. Using the above mentioned ab initio trajectories, we could evaluate corresponding correlation functions up to 2000 au with a time step of 20 au, which were then padded with zeros from 2000 au to 100000 au to obtain smooth spectra in the frequency domain. All single-Hessian thawed Gaussian correlation functions were computed using eq~\ref{eq:C_t_half_int}. In both harmonic and anharmonic calculations, time correlation functions were damped exponentially with lifetime parameters $\tau$ ranging from 800 to 3000 au (19--72 fs) (see Supporting Information). Computed spectra were shifted in frequency (see Supporting Information) to match the experiment at the highest intensity peak. All simulations were performed at zero temperature and within the Condon approximation, as discussed in Section~\ref{sec:theory}. For completeness, we tested for finite-temperature and non-Condon effects on the absorption spectrum of TFB by employing an extended version\cite{Patoz_Vanicek:2018,Begusic_Vanicek:2020,Begusic_Vanicek:2021a} of the single-Hessian thawed Gaussian approximation implemented in a separate code; results are presented in Section 6 of the Supporting Information.

\section{Results and discussion}
\subsection{\label{subsec:top_barrier}Initial wavepacket at the top of a potential energy barrier}
We first consider two examples in which the equilibrium geometry is of lower symmetry in the final state than in the initial electronic state. In such cases, two or more minima on the final-state potential energy surface are displaced along a non-totally symmetric mode. As a result, the initial wavepacket sits on top of a barrier connecting these minima. Within the adiabatic harmonic approach, this highly anharmonic potential energy surface is approximated by a harmonic one, centered at one of the potential energy wells, which is known to produce very inaccurate spectra.\cite{Begusic_Vanicek:2019} By imposing the Gaussian wavepacket ansatz, the thawed Gaussian dynamics also makes a crude approximation, since the actual nuclear wavepacket would eventually split and deform from its initial Gaussian shape. Nevertheless, if the wavepacket splitting occurs on a long time scale, more precisely, longer than the decay time of the correlation function, the features due to quantum-mechanical wavepacket splitting would not be visible in the spectrum and the crude semiclassical Gaussian wavepacket dynamics could yield a reasonable approximation of the spectrum.

\subsubsection{$\text{S}_1 \leftarrow \text{S}_0$ absorption spectrum of 1,2,4,5-tetrafluorobenzene}
Vibrationally resolved absorption and emission spectra of substituted fluorobenzene derivatives have been studied extensively in jet-cooled expansion experiments since the late 1970s and early 1980s.\cite{philis1981comparison, michaelahollas1993determination} Here, we focus on TFB, as one of the fluorobenzene derivatives exhibiting a double-well potential energy surface in the excited electronic state. Notably, while the ground-state equilibrium structure of TFB is planar with D$_{2\text{h}}$ symmetry, the excited-state minimum-energy structure is obtained by out-of plane bending of the fluorine atoms either above or below the original mirror plane, leading to two degenerate structures. The ``butterfly'' mode connecting these two minima was studied spectroscopically as a paradigmatic example of an anharmonic double-well system.\cite{orlandi1987vibronic,okuyama1986electronic}

\begin{figure}
    \centering
    \includegraphics{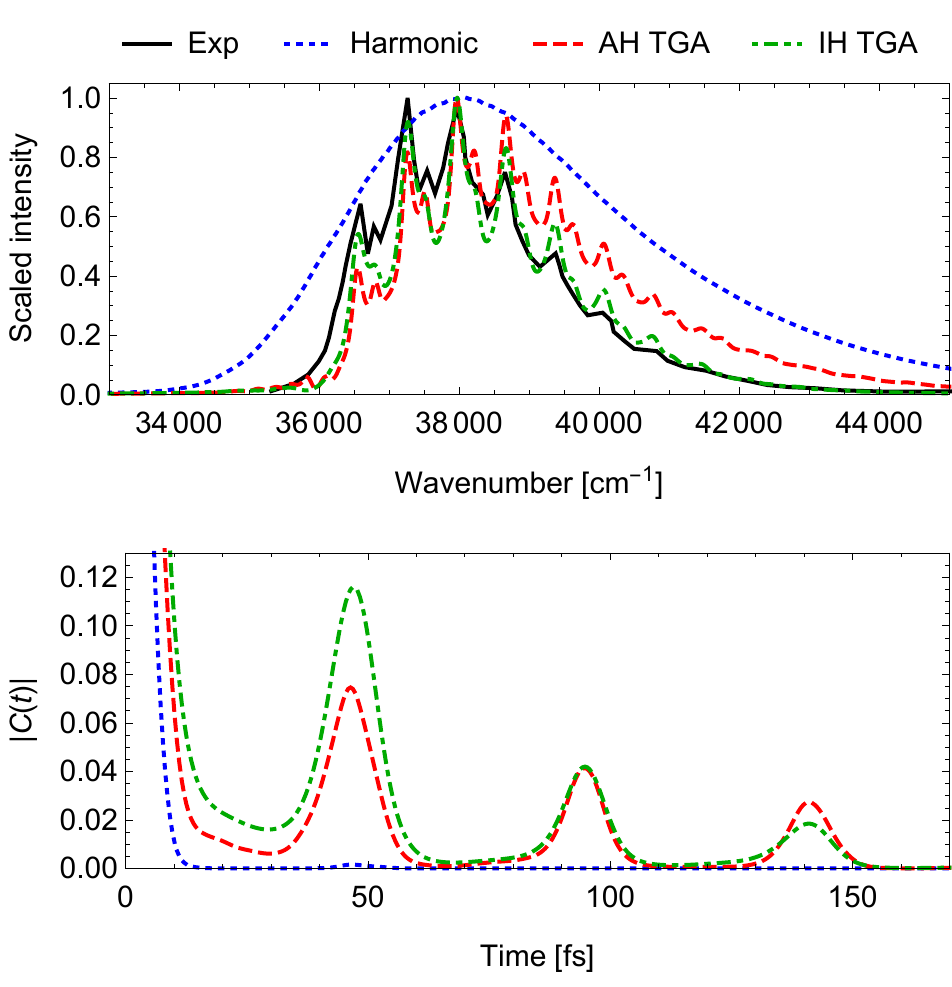}
    \caption{Top: $\text{S}_1 \leftarrow \text{S}_0$ absorption spectra of TFB computed using the adiabatic harmonic method (``Harmonic'') and single-Hessian thawed Gaussian approximation with either adiabatic Hessian (labeled ``AH TGA'') or initial (ground-state) Hessian (``IH TGA''), compared with the experimental spectrum recorded in the gas phase at room temperature.\cite{philis1981comparison} Bottom: Absolute value of the correlation function (first $170$~fs) computed using the harmonic method  (eq~\ref{eq:C_HA}) or the two variants of the single-Hessian thawed Gaussian approximation (both using eq~\ref{eq:C_t_half_int} and $\psi_{t}$ propagated with eqs~\ref{eq:psi_dot} and \ref{eq:V_SHA}).}
    \label{fig:TFB}
\end{figure}

The anharmonic simulations (Figure~\ref{fig:TFB}, top), based on either adiabatic or initial single-Hessian thawed Gaussian approximations, of the absorption spectrum of TFB correctly reproduce the double-headed progression found in the experiment. In contrast, the harmonic spectrum exhibits no vibronic structure due to a quick decay of the wavepacket autocorrelation function (see Figure~\ref{fig:TFB}, bottom). This initial damping of the correlation function occurs irrespective of the chosen lifetime parameter $\tau$; it occurs due to dynamics along the symmetry-breaking mode, which is inadequately modeled as a highly displaced harmonic oscillator.

The initial-Hessian thawed Gaussian approximation performs somewhat better than the adiabatic-Hessian approach, a result that was also observed in quinquethiophene emission spectrum in ref~\citenum{Begusic_Vanicek:2019}. This is not directly related to the large displacement of the excited-state geometry along the non-symmetric mode, which is not used in the thawed Gaussian wavepacket simulations. Rather, the Hessian evaluated at the displaced excited-state minimum is inadequate to approximate locally the potential energy surface about the center of the wavepacket. When the ground-state Hessian is employed, the width of the wavepacket is constant, which can yield a better approximation to the exact wavepacket dynamics, as is the case here. Surprisingly, the vertical-Hessian method performs worse (see Supporting Information, Section 7, Figure S5) and misses the experimentally observed vibronic progression, much like the harmonic approximation. Because the vertical Hessian exhibits three imaginary frequencies (with magnitudes $421$, $350$, and $191$ cm$^{-1}$, see Section~3.4 of the Supporting Information), the Gaussian wavepacket propagated using this Hessian spreads quickly, resulting in a fast decay of the wavepacket autocorrelation function and an incorrectly broadened spectrum.

For completeness, in Section 6  of the Supporting Information, we computed the spectra of TFB also at finite temperature (Figure S3) and beyond the Condon approximation (Figure S4). The inclusion of finite-temperature effects improves slightly the (more accurate) spectrum computed with the initial Hessian, but over-broadens the (less accurate) spectrum computed with the adiabatic Hessian. The non-Condon effects are completely negligible in both cases, partially because the Condon approximation is unaffected by the variation of the transition dipole moment along the trajectory and requires only that this moment be approximately constant in the region of the initial state.

\subsubsection{Ammonia $\text{S}_1 \rightarrow \text{S}_0$ emission}

\begin{figure}
    \centering
    \includegraphics{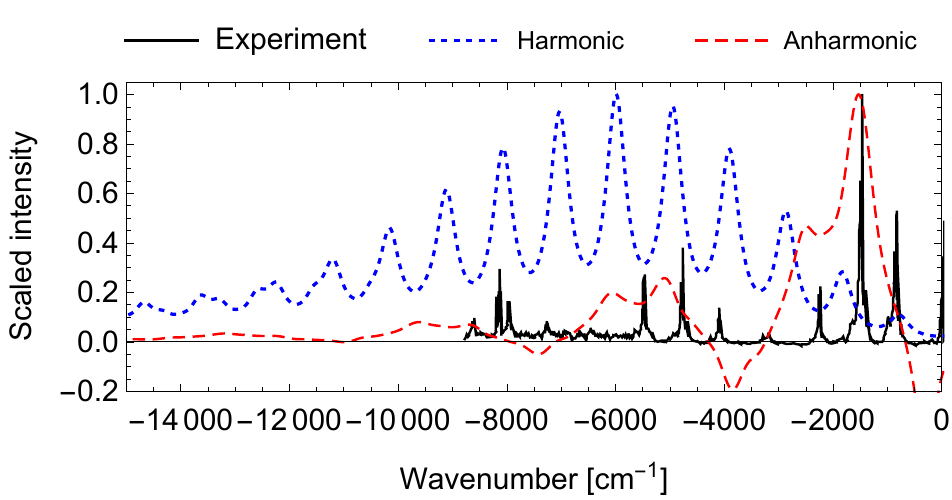}
    \caption{$\text{S}_1 \rightarrow \text{S}_0$ emission spectra of ammonia computed using the adiabatic harmonic method (``Harmonic'') and adiabatic single-Hessian thawed Gaussian approximation (``Anharmonic''), compared with the experimental spectrum recorded in the gas phase.\cite{Tang_Imre:1991}}
    \label{fig:NH3_Emission_Spectrum}
\end{figure}

The experimental emission spectrum of ammonia (Figure~\ref{fig:NH3_Emission_Spectrum}) clearly shows two progressions: a high-frequency progression corresponding to the bond-stretch mode, and a low-frequency progression due to the excitation of the umbrella-inversion mode. Although simple, the spectrum is determined by a rich wavepacket dynamics in the ground electronic state. Tang, Abramson, and Imre\cite{Tang_Imre:1991} used a two-dimensional model to propagate exactly the nuclear wavepacket and showed that wavepacket splitting and delocalization are inevitable processes in ammonia fluorescence. The single-Hessian thawed Gaussian method cannot describe these effects and, therefore, captures only the high-frequency progression due to the symmetric stretching mode. In contrast to our previous example, here the wavepacket splitting effects appear before the correlation function is damped by external broadening effects and contribute to the spectrum. The adiabatic harmonic model, which approximates the ground-state potential energy surface as a single harmonic well, produces an artificial and incorrect long progression. In this case, both single-Hessian thawed Gaussian and harmonic approximations are inadequate.

\subsection{\label{subsec:displaced_barrier}Umbrella mode in ammonia, phospine, and arsine}

In Section~\ref{subsec:top_barrier} we discussed examples in which the wavepacket is initially at the top of a barrier of a double-well potential. In this Section, we investigate cases in which the initial wavepacket is displaced and approaches the barrier only at a later time. This happens, for example, in the photoelectron spectra of phosphine and arsine. For comparison, we also present the results for ammonia.

\begin{figure}
    \centering
    \includegraphics[scale=1]{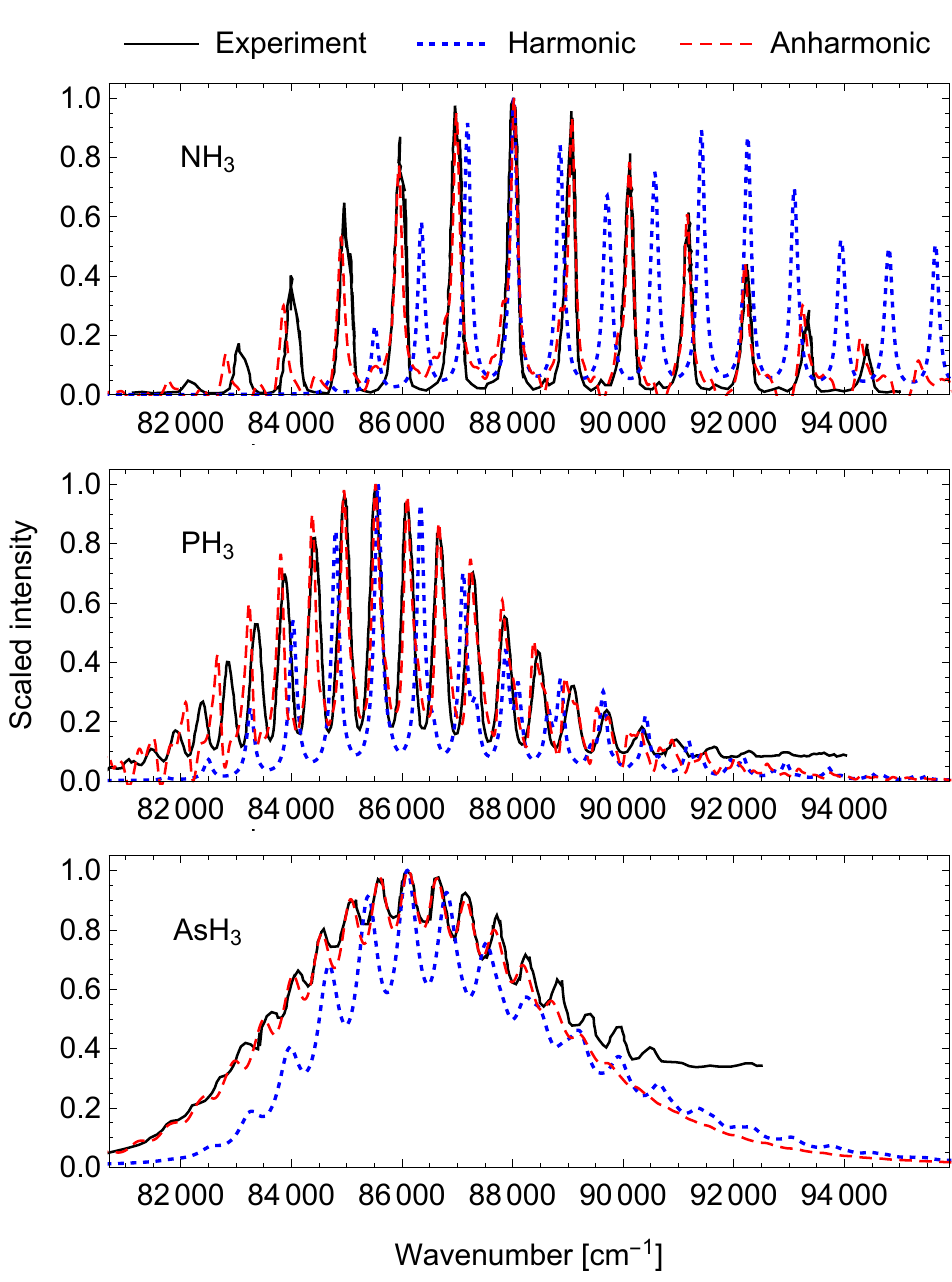}
    \caption{Experimental, adiabatic harmonic, and adiabatic single-Hessian thawed Gaussian (``Anharmonic'') photoelectron spectra of ammonia (top),\cite{Edvardsson_Rennie:1999} phosphine (middle),\cite{maripuu1983hei} and arsine (bottom).\cite{Branton_Stenhouse:1970} Spectra were aligned at the peak with maximum intensity.}
    \label{fig:SpectraPart2}
\end{figure}

Photoelectron spectrum of ammonia has been used extensively as a test case for methods that treat anharmonicity effects on vibrationally resolved spectra. Because the equilibrium geometry in the cationic state is planar (D$_{3\text{h}}$ symmetry), the nuclear dynamics following electron ejection is characterized by a large-amplitude umbrella motion, leading to a long progression in the photoelectron spectrum. Although the potential energy surface of the cationic state can be charaterized as a single-well potential, it is poorly represented by a quadratic function. Indeed, the harmonic approach (Figure~\ref{fig:SpectraPart2}, top) fails to reproduce both spacings and intensities of the experimental spectrum. It was shown in ref~\citenum{Wehrle_Vanicek:2015} that the thawed Gaussian approximation can largely improve on the harmonic result. In Figure~\ref{fig:SpectraPart2} (top), we demonstrate that this holds even for its single-Hessian version.

\begin{figure}
    \centering
    \includegraphics[scale=0.4]{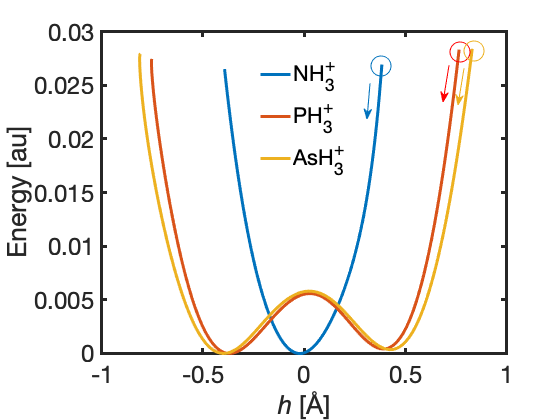}
    \caption{Potential energy as a function of the signed distance $h$ between atom X (X=N, P, As) and the center of mass of the three hydrogen atoms, obtained from the first 65, 120, and 135 molecular dynamics steps, respectively, after vertical ionization to NH$_3^+$, PH$_3^+$, and AsH$_3^+$. The lowest energy value was set to zero in each case. The Franck--Condon points and the direction of the initial dynamics are indicated by the circles and arrows, respectively.}
    \label{fig:PES_comparison}
\end{figure}

Interestingly, phosphine and arsine, whose ground-state structure is similar to ammonia, are not planar in the cationic state. In contrast to ammonia, their equilibrium geometries in the cationic state are bent along the umbrella mode, while the planar geometries are barriers connecting the energy minima. Nevertheless, the barriers in PH$_3^{+}$ and AsH$_3^{+}$ are much lower in energy than the Franck--Condon point, which is why the nuclear dynamics can cross over the barrier in a classical fashion (see Figure~\ref{fig:PES_comparison}), justifying the semiclassical thawed Gaussian dynamics picture. Indeed, the middle and bottom panels of Figure~\ref{fig:SpectraPart2} demonstrate that the crude thawed Gaussian spectra excellently reproduce the main features found in the experimental spectra. In contrast, the harmonic approximation overestimates the frequency spacings of the vibronic progressions and fails to recover the spectral peak intensities.

Notably, the harmonic approximation overestimates the frequency of the umbrella mode in PH$_3$ and AsH$_3$, but underestimates it in NH$_3$. Due to the double-well potential energy surfaces of phosphine and arsine cations, the harmonic potential is constructed in one of the two wells and necessarily assumes a shorter period of oscillation. In contrast, the harmonic potential constructed about the single minimum of the ammonium cation potential energy surface is less stiff along the umbrella mode, leading to a lower effective frequency.

Finally, we note that the inaccuracy of the single-Hessian thawed Gaussian spectra is pronounced especially in the low-frequency region of the phosphine photoelectron spectrum. This is explained by the inability of the single-Gaussian wavepacket dynamics to describe scattering off the barrier walls or tunneling, which would be needed to account for the double-well structure of the eigenstates that are beneath or just above the potential barrier. In arsine, this effect is less prominent due to the low resolution of the photoelectron spectrum.

\section{Conclusion}

By studying various representative examples, we hope to have provided guidelines for using the recently proposed single-Hessian thawed Gaussian approximation in systems where the conventional harmonic approximation fails. Although approximate, the thawed Gaussian method can reproduce experimental vibrationally resolved electronic spectra even when the potential energy surfaces are fairly anharmonic and exhibit double-well character. Yet, this is subject to restrictions, as shown in the high-resolution emission spectrum of ammonia. As a rule of thumb, the single-Hessian thawed Gaussian approximation is expected to yield reasonable results in the following two cases that are challenging for the harmonic approximation:

(i) When the Franck--Condon point is at the top of a potential energy barrier along a non-symmetric mode, provided that the spectrum is sufficiently broad, i.e., that the wavepacket splitting occurs only after the time correlation function is damped.

(ii) When the potential energy barrier is very small compared to the initial potential energy at the Franck--Condon geometry, which allows for the wavepacket to move across the barrier in a classical fashion.

We have not found a simple way to use the local features of the potential energy surfaces, such as the optimized geometries and corresponding Hessians (i.e., the data needed for evaluating the harmonic spectra), for predicting the accuracy of the single Hessian approximation.

Apart from these practical considerations, we presented and employed a new robust implementation of the single-Hessian thawed Gaussian wavepacket dynamics. The method will soon be available in the widely used TURBOMOLE quantum chemistry package, allowing the efficient calculation of anharmonic spectra using a variety of ab initio methods. The thawed Gaussian approach could serve as an alternative to conventional harmonic methods, allowing to identify the degree of anharmonicity effects on vibrationally resolved absorption, emission, and photoelectron spectra.

\begin{acknowledgement}
T.B. and J.V. acknowledge the financial support from the European Research Council (ERC) under the European Union's Horizon 2020 research and innovation programme (grant agreement No. 683069 -- MOLEQULE). E.T. acknowledges support of the National Institutes of Health (NIH) under award number R15GM126524. The content is solely the responsibility of the authors and does not necessarily represent the official views of the NIH. We also acknowledge technical support from the Division of Information Technology of CSULB.
\end{acknowledgement}

\begin{suppinfo}
Derivation of eq~\ref{eq:alpha_t_det_sqrt}, optimized geometries, harmonic frequencies, comparison between eqs~\ref{eq:C_t} and \ref{eq:C_t_half_int} for computing the thawed Gaussian wavepacket autocorrelation function, broadening parameters and frequency shifts applied to computed spectra, analysis of finite-temperature and non-Condon effects on the TFB absorption spectrum, vertical-Hessian thawed Gaussian simulation of the TFB spectrum.
\end{suppinfo}

\bibliography{biblio52,additions_DoubleWell}

\end{document}

\section{Computing the square root of the determinant of a complex symmetric matrix with positive definite real part}

Let $M_t$ be a complex symmetric $D \times D$ matrix with positive definite real part and let $M_t$ be a continuous function of a real parameter (time) $t$. We wish to compute
\begin{equation}
    f(t) = (\det M_t)^{1/2},
\end{equation}
such that $f(t)$ is also continuous in time. This is a non-trivial problem because the principal complex square root, defined as
\begin{equation}
    \sqrt{z} = \sqrt{|z|}e^{i\phi/2}, \phi = \arg z, -\pi < \phi < \pi,
\end{equation}
is discontinuous along the nonpositive real axis (branch cut). Therefore,
\begin{equation}
    f(t) = \sqrt{\det M_t}
\end{equation}
is not continuous in time. We will show below that
\begin{equation}
    f(t) = \prod_{j=1}^{D} \sqrt{\lambda_{j,t}}, \label{eq: f_t_cont}
\end{equation}
where $\lambda_{i,t}$ are the eigenvalues of $M_t$, is a continuous function of time.

First, we consider a complex scalar function of time, $a(t)$, restricted to $\text{Re}[a(t)] > 0$ for all times $t$. Then, its principal square root is a continuous function of time because $a(t)$ never crosses the branch cut.

Second, we prove that the eigenvalues of a complex symmetric $D \times D$ matrix $M = A + iB$ with a positive definite real part $A$ satisfy
\begin{equation}
    \text{Re} \lambda_j > 0.
\end{equation}
\textit{Proof:} For a complex vector $v$, we have
\begin{align}
v^{\dagger} \cdot M \cdot v &= v^{\dagger} \cdot A \cdot v + i v^{\dagger} \cdot B \cdot v \label{eq:M_v_squared_1}\\
                            &= \sum_j \frac{1}{\lVert a_j \rVert^2} (v^{\dagger} \cdot A \cdot a_j) (a_j^{\dagger} \cdot v) + i \sum_j \frac{1}{\lVert b_j \rVert^2} (v^{\dagger} \cdot b_j) (b_j^{\dagger} \cdot B \cdot v) \label{eq:M_v_squared_2}\\
                            &= \sum_j \lambda_{A, j}\frac{|\alpha_{j}|^2}{\lVert a_j \rVert^2} + i \sum_j \lambda_{B, j}\frac{|\beta_{j}|^2}{\lVert b_j \rVert^2}, \label{eq:M_v_squared_3}
\end{align}
where $a_j$ ($b_j$) are the column eigenvectors of matrix $A$ ($B$), $\lambda_{A, j}$ ($\lambda_{B, j}$) are the associated real eigenvalues, and
\begin{align}
\alpha_j &= a_j^{\dagger} \cdot v \\
\beta_j &= b_j^{\dagger} \cdot v.
\end{align}
In going from eq~\ref{eq:M_v_squared_1} to \ref{eq:M_v_squared_2}, we used the fact that all real symmetric matrices can be diagonalized by an orthogonal transformation. From eq~\ref{eq:M_v_squared_3} we conclude that 
\begin{equation}
\text{Re}(v^{\dagger} \cdot M \cdot v) > 0, \label{eq:real_exp_val}
\end{equation} 
because all eigenvalues $\lambda_{A, j}$ and $\lambda_{B, j}$ are real and $\lambda_{A,j} > 0, \forall~j$.
Now let $v$ be one of the eigenvectors of $M$ and $\lambda$ the corresponding eigenvalue. Then, from
\begin{equation}
v^{\dagger} \cdot M \cdot v = \lambda v^{\dagger} \cdot v
\end{equation}
and eq~\ref{eq:real_exp_val}, it follows that
\begin{equation}
\text{Re}\lambda = \frac{\text{Re}(v^{\dagger} \cdot M \cdot v)}{\Vert v \Vert^2} > 0,
\end{equation}
which completes the proof.

Third, we note that the eigenvalues $\lambda_{j=1,\dots,D}$ of a $D \times D$ matrix $M$ are roots of the characteristic polynomial
\begin{equation}
    p(\lambda) = \det(M - \lambda I_D) = \prod_{j=1}^{D} (\lambda_j - \lambda).
\end{equation}
It follows that the determinant of a matrix can always be computed as
\begin{equation}
    p(0) = \det M = \prod_{j=1}^{D} \lambda_j,
\end{equation}
even if $M$ is non-diagonalizable.

To summarize, evaluating $f(t)$ according to eq~\ref{eq: f_t_cont} yields a continuous function of time because the eigenvalues of $M_t$ have positive real parts and their principal square root is continuous.

\section{Cartesian coordinates of the equilibrium structures in Bohr}
\subsection{NH$_3$ (CC2/aug-cc-pVTZ)}\footnotesize
\begin{Verbatim}[baselinestretch=0.75]
neutral ground state
    0.83130330496479     -0.59150330792531      1.56581657416454  h
    0.93963442263971     -0.59152254339342     -1.50327013645619  h
   -1.77243993297970     -0.59110491039056     -0.06254423529471  h
   -0.00038750735279      0.13197040092456     -0.00000220241349  n
charged ground state
    0.90413857634592     -0.41067733798368      1.70287090988172  h
    1.02195164914944     -0.41069951301932     -1.63485142435267  h
   -1.92750733819904     -0.41024357650236     -0.06801962855033  h
   -0.00047260002437     -0.41053993331812      0.00000014302138  n
first excited singlet state
   0.92535994795955     -0.41185001190813      1.74280707179170    h
    1.04593813368959     -0.41179070373000     -1.67320525334941   h
   -1.97270379719945     -0.41134011182828     -0.06962055546265   h
   -0.00046811389092     -0.41165127159525     -0.00000500199662   n
\end{Verbatim}
\subsection{PH$_3$ (CC2/aug-cc-pVTZ)}\footnotesize
\begin{Verbatim}[baselinestretch=0.75]
neutral ground state
    1.05348500100859     -0.77152937972078      1.98411198081460  h
    1.19075655081793     -0.77155523307171     -1.90485816870274  h
   -2.24582511792627     -0.77102391892551     -0.07925255116557  h
   -0.00030614667520      0.67194813016380     -0.00000126094626  p
charged ground state
    1.19188557671503     -0.60101674397154      2.24363702708525  h
    1.34793532303470     -0.57349508375895     -2.15576892634737  h
   -2.54126654344968     -0.58384879526082     -0.09114532900110  h
   -0.00044406907783      0.11620021775647      0.00327722826424  p
\end{Verbatim}
\subsection{AsH$_3$ (CC2/aug-cc-pVTZ)}\footnotesize
\begin{Verbatim}[baselinestretch=0.75]
neutral ground state
    0.00539513327242     -0.00247016516622      1.17417061065099  as
   -0.41795939666559      2.32604447674264     -0.38458512496632  h
   -1.80740620577313     -1.52221781133969     -0.38629991733673  h
    2.21997046916633     -0.80135650023667     -0.40328556834786  h
charged ground state
    0.00264243720921     -0.00120993855057      0.57513184675463  as
   -0.47302341441808      2.63840060134778     -0.18399017528702  h
   -2.04936616814002     -1.72750718606242     -0.18593548222857  h
    2.51974714534884     -0.90968347673478     -0.20520618923898  h
\end{Verbatim}
\subsection{1,2,4,5-Tetrafluorobenzene (CC2/def2-TZVP)}\footnotesize
\begin{Verbatim}[baselinestretch=0.75]
neutral ground state
   -4.46354111046531     -2.56255882785211      0.00000000000000  f
   -4.46354111046531      2.56255882785211      0.00000000000000  f
    4.46354111046531      2.56255882785211      0.00000000000000  f
    4.46354111046531     -2.56255882785211      0.00000000000000  f
   -2.25926821719139      1.31478228907706      0.00000000000000  c
   -2.25926821719139     -1.31478228907706      0.00000000000000  c
    2.25926821719139     -1.31478228907706      0.00000000000000  c
    2.25926821719139      1.31478228907706      0.00000000000000  c
    0.00000000000000     -2.64520208188283      0.00000000000000  c
    0.00000000000000      2.64520208188283      0.00000000000000  c
    0.00000000000000     -4.69181724370117      0.00000000000000  h
    0.00000000000000      4.69181724370117      0.00000000000000  h
first excited singlet state
   -4.40712848687314     -2.49158603520767      0.27758567173279  f
   -4.40725111129323      2.48570864928358      0.40413040490876  f
    4.40726735396786      2.48567738473168      0.40414492257560  f
    4.40711173743581     -2.49161698378589      0.27760065180757  f
   -2.21468660242481      1.35436320599405     -0.14969993615190  c
   -2.21470072889866     -1.33325501131817     -0.21778170823795  c
    2.21469031991536     -1.33327070345082     -0.21777061536444  c
    2.21469596941645      1.35434544484468     -0.14968966370205  c
   -0.00000783647665     -2.78880023431227     -0.56005504452959  c
    0.00001081243573      2.82532873716004     -0.41957428254920  c
   -0.00001639964206     -4.74352072814417     -1.16078029416937  h
    0.00002071243721      4.80814638420356     -0.91996138628469  h
\end{Verbatim}
\newpage

\section{Harmonic frequencies}
\begin{Verbatim}[baselinestretch=0.75]
#  mode     symmetry     wave number   IR intensity  
#                         cm**(-1)        km/mol  
\end{Verbatim}
\subsection{NH$_3$ (CC2/aug-cc-pVTZ)}\footnotesize
\begin{Verbatim}[baselinestretch=0.75]
neutral ground state
     7        a            1041.16       132.74740  
     8        a            1665.14        13.25284   
     9        a            1665.58        13.24905    
    10        a            3485.35         4.53737     
    11        a            3628.96         6.32081      
    12        a            3629.30         6.31865       
charged ground state
     7        a             835.41       221.05866      
     8        a            1555.36        65.28629      
     9        a            1555.94        65.29942      
    10        a            3397.10         0.00091      
    11        a            3587.70       293.24965      
    12        a            3588.16       293.21165      
first excited singlet state
     7        a             729.79        96.80586      
     8        a            1354.80       368.20317      
     9        a            1355.25       367.90156      
    10        a            2930.70         0.01348      
    11        a            3101.81      2441.42692      
    12        a            3102.26      2439.90411      
\end{Verbatim}
\subsection{PH$_3$ (CC2/aug-cc-pVTZ)}\footnotesize
\begin{Verbatim}[baselinestretch=0.75]
neutral ground state
     7        a            1009.87        20.74487      
     8        a            1155.16        13.87668      
     9        a            1155.20        13.88857      
    10        a            2463.46        28.70584      
    11        a            2474.92        52.45561      
    12        a            2475.07        52.41385      
charged ground state
     7        a             768.80         1.15467      
     8        a            1067.22         8.69011      
     9        a            1067.51         8.68524      
    10        a            2533.48        12.63522      
    11        a            2607.40        54.28874      
    12        a            2607.90        54.51698      
\end{Verbatim}

\subsection{AsH$_3$ (CC2/aug-cc-pVTZ)}\footnotesize
\begin{Verbatim}[baselinestretch=0.75]
neutral ground state
     7        a             930.45        28.65107       
     8        a            1047.32        14.82433       
     9        a            1047.35        14.72368       
    10        a            2307.85        61.07369       
    11        a            2340.69        90.37984       
    12        a            2340.77        90.15764       
charged ground state
     7        a             710.64         1.29383       
     8        a             960.65         7.03396       
     9        a             961.33         7.04069       
    10        a            2415.10         8.67950       
    11        a            2483.39        33.10605       
    12        a            2483.58        33.08328       
\end{Verbatim}
\subsection{1,2,4,5-Tetrafluorobenzene (CC2/def2-TZVP)}\footnotesize
\begin{Verbatim}[baselinestretch=0.75]
neutral ground state
     7        a             127.98         0.00000       
     8        a             197.02         1.66482       
     9        a             271.48         0.00000       
    10        a             286.35         0.00000       
    11        a             293.56         0.18031       
    12        a             342.42         3.18263       
    13        a             413.54         0.00000       
    14        a             417.57         0.00000       
    15        a             465.17         4.56955       
    16        a             486.32         0.00000       
    17        a             601.79         0.00000       
    18        a             632.78         0.00000       
    19        a             667.22         0.00000       
    20        a             700.91        27.20669       
    21        a             750.38         0.00000       
    22        a             804.54         0.00000       
    23        a             865.71       100.18778       
    24        a             870.30        58.86356       
    25        a            1138.45         0.00000       
    26        a            1179.31       128.23039       
    27        a            1208.47         0.00000       
    28        a            1219.10       115.39424       
    29        a            1401.81         0.00000       
    30        a            1421.16        25.47791       
    31        a            1450.17        41.42818       
    32        a            1555.63       471.70536       
    33        a            1657.64         0.00000       
    34        a            1662.10         0.00000       
    35        a            3234.01        11.51424       
    36        a            3235.62         0.00000       
first excited singlet state optimized geometry
     7        a              90.56         0.01027       
     8        a              97.29         0.00000       
     9        a             131.94         0.00154       
    10        a             267.90         0.76700       
    11        a             297.83         0.68996       
    12        a             320.53        22.61356       
    13        a             362.18         4.92400       
    14        a             370.56         0.00004       
    15        a             388.31         0.80777       
    16        a             407.13        80.39501       
    17        a             429.95         5.04035       
    18        a             503.09         1.23665       
    19        a             539.10        73.71195       
    20        a             561.36         0.00000       
    21        a             604.60         0.00000       
    22        a             665.13        23.78849       
    23        a             724.21         1.83518       
    24        a             809.98        54.42942       
    25        a            1098.05         0.00000       
    26        a            1153.24       249.64267       
    27        a            1160.26       140.85516       
    28        a            1161.01         0.00006       
    29        a            1331.11         9.12037       
    30        a            1365.63        30.99582       
    31        a            1393.74       807.39442       
    32        a            1495.26         0.00040       
    33        a            1570.67         5.60741       
    34        a            1708.04       259.43777       
    35        a            3221.62         0.00075       
    36        a            3224.66         0.11261    
first excited singlet state (Franck-Condon geometry)
     7        a            -421.25         0.00000 
     8        a            -350.80         0.00000 
     9        a            -191.60         0.00000 
    10        a              98.31         0.00000 
    11        a             180.77         0.00000 
    12        a             236.95         0.00000 
    13        a             257.64         0.00000 
    14        a             279.06         6.73582 
    15        a             300.02         1.63135 
    16        a             332.10         0.00000 
    17        a             342.35         3.01048 
    18        a             361.83         0.00000 
    19        a             396.18         0.00000 
    20        a             423.17         0.00000 
    21        a             615.51         0.00000 
    22        a             693.50        17.08683 
    23        a             741.71         0.00000 
    24        a             844.78        67.36671 
    25        a            1128.92         0.00000 
    26        a            1183.12       155.28389 
    27        a            1183.31       161.77025 
    28        a            1216.02         0.00000 
    29        a            1403.88        14.61086 
    30        a            1412.90         0.00000 
    31        a            1515.56       575.19198 
    32        a            1661.68         0.00000 
    33        a            1695.83         0.00000 
    34        a            1853.86       191.79028 
    35        a            3230.79        16.71346 
    36        a            3232.45         0.00000 
\end{Verbatim}  

\section{Comparison between the original method for computing the correlation function and the approach using the half-time wavepacket}

\begin{figure}[H]
    \centering
    \includegraphics{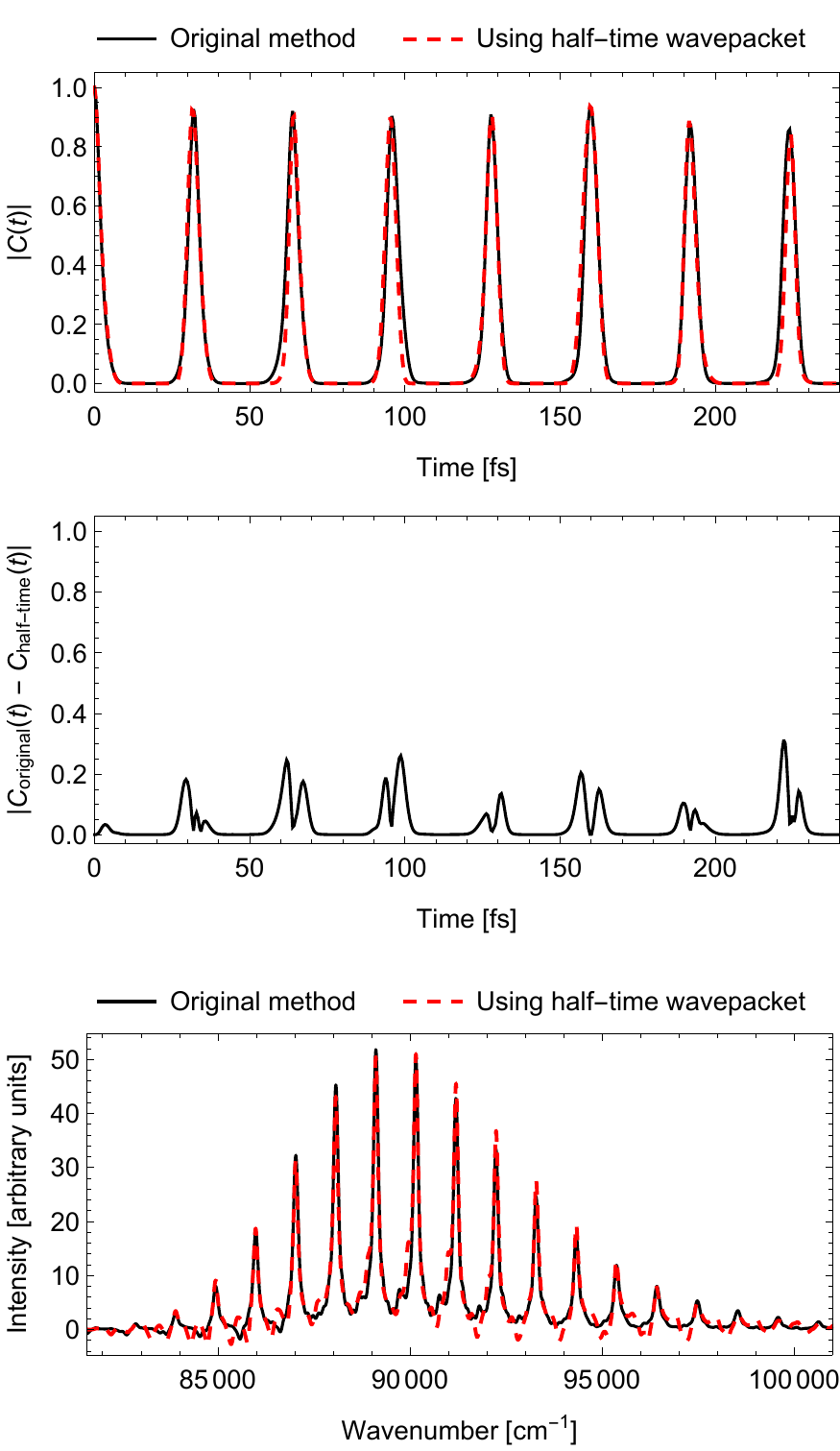}
    \caption{Top: Absolute values of the correlation functions evaluated with eq~2 (black) and eq~23 (red, dashed) of the main text. Correlation functions correspond to the photoelectron spectrum of ammonia discussed in Section~4.2 of the main text. Middle: Absolute value of the difference between the two correlation functions. Bottom: Corresponding spectra. In this example, the spectra computed using the half-time approach (eq~23 of the main text) exhibit more negative spectral intensities than the spectra computed with eq~2. However, opposite behavior can also be observed, as shown on a model system below (Figure~\ref{fig:quartic}).}
    \label{fig:halftime}
\end{figure}

\begin{figure}[H]
    \centering
    \includegraphics{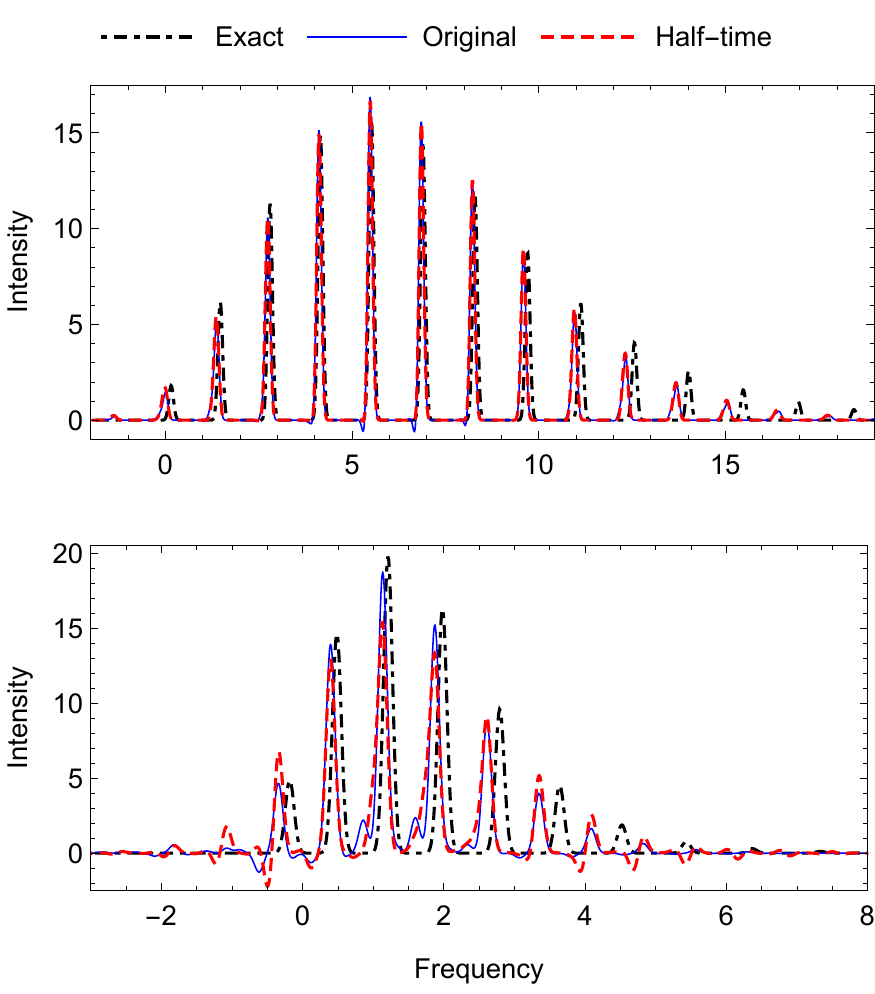}
    \caption{Exact and adiabatic single-Hessian thawed Gaussian absorption spectra of two model systems composed of a harmonic ground-state potential $V(q) = 0.5 \, q^2$, which sets the initial Gaussian wavepacket parameters to $q_0 = 0$, $p_0 = 0$, $Q_0 = 1$, $P_0 =i$, with $m=\hbar=1$, and a quartic excited-state potential $V(q) = 0.5 \, \omega_{e}^{2} q^2 + 0.01 \, q^4$, where $\omega_{e} = 1.3$ (top panel) or $\omega_{e} = 0.6$ (bottom). In the top panel, the spectrum computed using the original wavepacket autocorrelation function expression (eq~2 of the main text) exhibits more negative intensity, whereas in the bottom panel it is the half-time wavefunction approach (eq~23 of the main text) that results in more negative spectral features. These negative intensities are a consequence of the local potential approximation used for propagating the thawed Gaussian wavepacket, as discussed in a previous work.\cite{Begusic_Vanicek:2019}}
    \label{fig:quartic}
\end{figure}

\section{Broadening parameters and frequency shifts applied to computed spectra}

\begin{table}[H]
    \centering
    \begin{tabular}{lr}
     \hline
     Spectrum & $\tau$ / atomic units \\ \hline
     TFB absorption   $\qquad$ & 2200 \\
     NH$_3$ emission           & 800 \\
     NH$_3$ photoelectron      & 3000 \\
     PH$_3$ photoelectron      & 2700 \\
     AsH$_3$ photoelectron     & 1000 \\\hline
    \end{tabular}
    \caption{Lifetimes $\tau$ of the exponential function used for damping the correlation functions.}
    \label{tab:lifetimes}
\end{table}

\begin{table}
    \centering
    \begin{tabular}{lrrrr}
     \hline
                          & Harmonic & AH TGA & IH TGA & VH TGA \\ \hline
     TFB absorption       & -3600    & -1200  & -2250  & -1800 \\
     NH$_3$ emission      & -44500   & -34000 &        &  \\
     NH$_3$ photoelectron &  2200    &  80    &        &  \\
     PH$_3$ photoelectron &  3600    &  1450  &        &  \\
     AsH$_3$ photoelectron& 4550     & 1550  &        &  \\\hline
    \end{tabular}
    \caption{Frequency shifts applied to the computed spectra to match them to the experiment, as discussed in the main text. NH$_3$ emission spectra were available on a relative frequency scale only (see Figure 2 of the main text), so the shifts reported here can be used to obtain the absolute frequency scale of the computed spectra.}
    \label{tab:shifts}
\end{table}
\newpage
\section{Finite-temperature and non-Condon effects on the TFB absorption spectrum}

\begin{figure}
    \centering
    \includegraphics{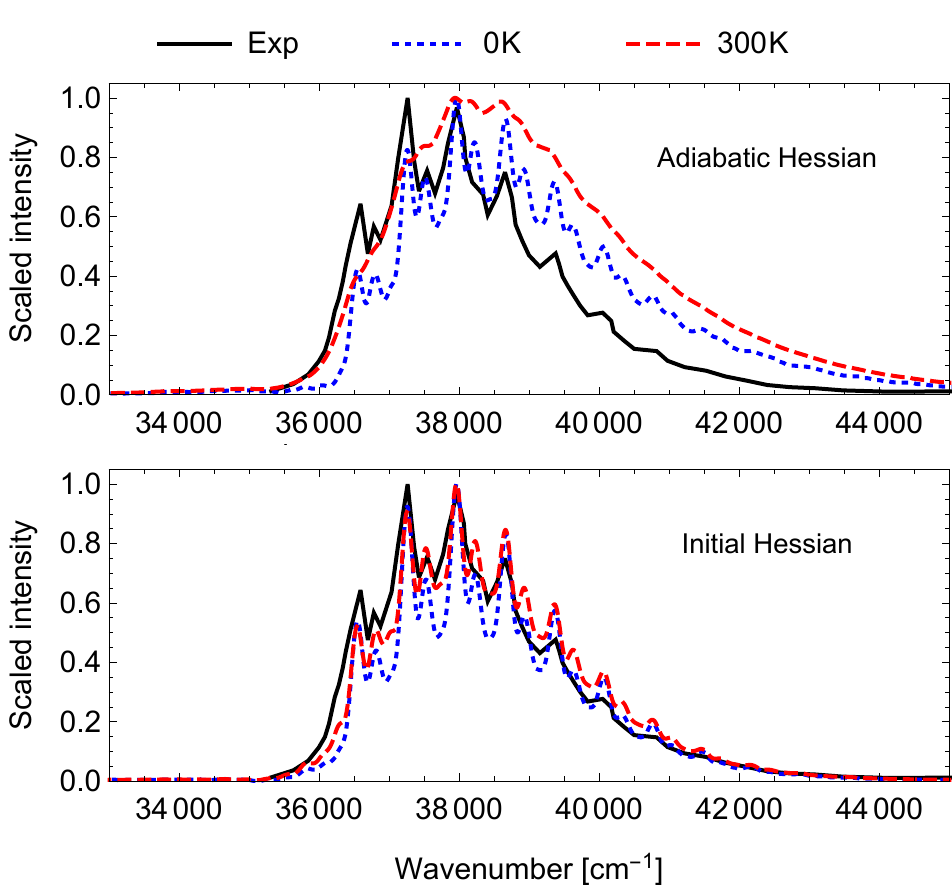}
    \caption{$\text{S}_1 \leftarrow \text{S}_0$ absorption spectrum of TFB computed using the adiabatic-Hessian (top) or initial-Hessian (bottom) thawed Gaussian approximation at $0\ $K (blue, dotted) or $300\ $K (red, dashed), compared with the experimental spectrum (black). Temperature effects are responsible for the broadening of the spectrum, especially when the adiabatic Hessian is used. The finite-temperature spectrum was computed within the Condon approximation, using the method outlined in Ref.~\citenum{Begusic_Vanicek:2020}.}
    \label{fig:TFB_temperature}
\end{figure}

\begin{figure}
    \centering
    \includegraphics{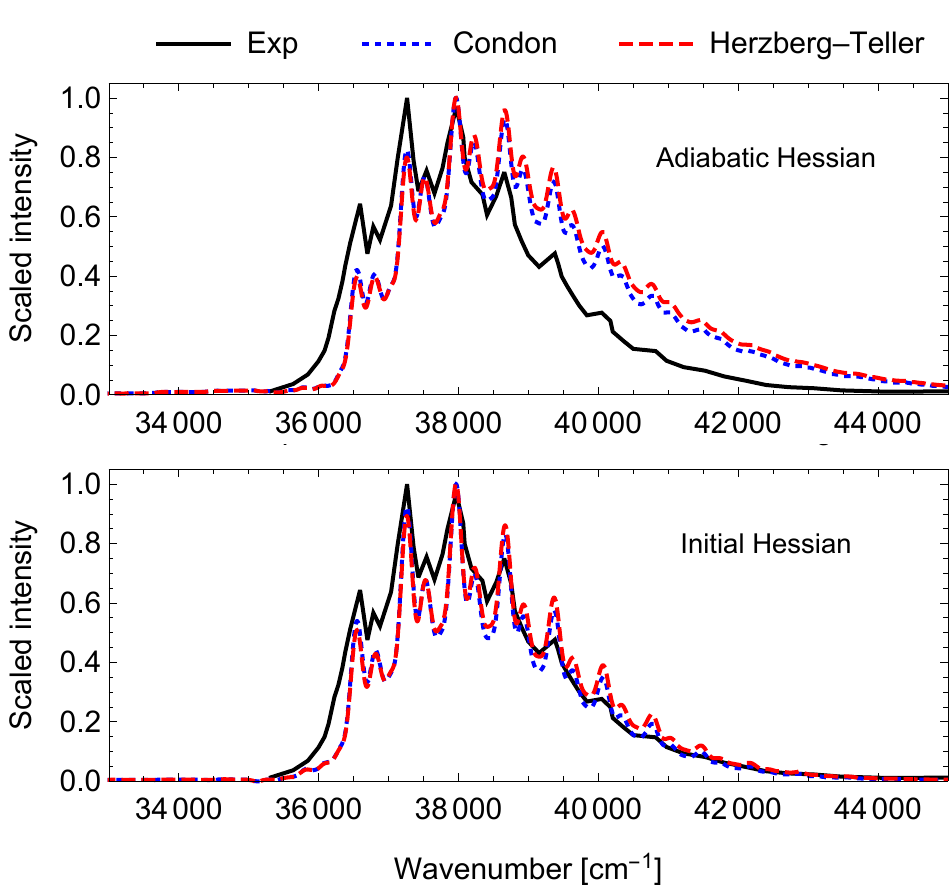}
    \caption{$\text{S}_1 \leftarrow \text{S}_0$ absorption spectrum of TFB computed using the adiabatic-Hessian (top) or initial-Hessian (bottom) thawed Gaussian approximation within Condon (blue, dotted) or Herzberg--Teller (red, dashed) approximations of the transition dipole moment, compared with the experimental spectrum (black). The results show that the non-Condon effects are negligible. Spectra were computed at zero temperature, using the method outlined in Refs.~\citenum{Patoz_Vanicek:2018} and \citenum{Prlj_Vanicek:2020}. For the Herzberg--Teller contribution to the spectrum, we numerically (nuclear step 0.02\,a.u.) computed the gradient of the (right) CC2 transition dipole moment at the ground-state optimized geometry.}
    \label{fig:TFB_FCHT}
\end{figure}

\newpage
\section{Vertical-Hessian thawed Gaussian simulation of the TFB absorption spectrum}

\begin{figure}
    \centering
    \includegraphics{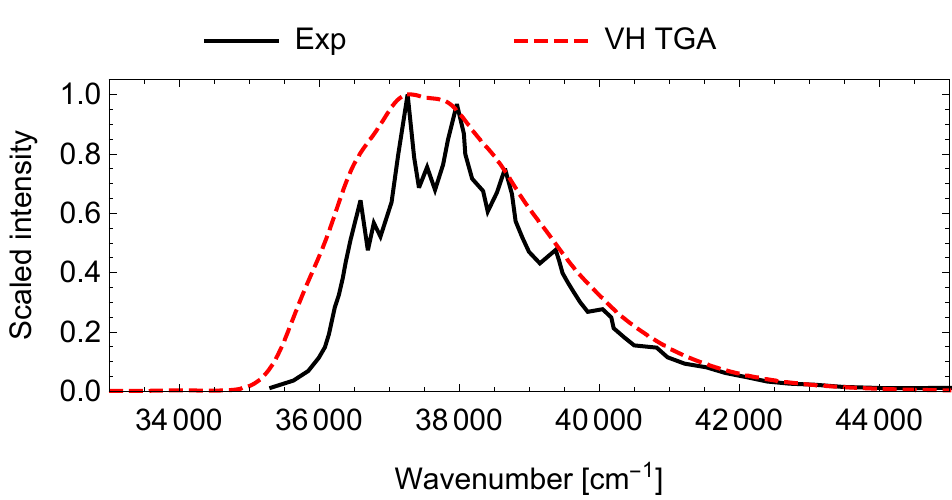}
    \caption{$\text{S}_1 \leftarrow \text{S}_0$ absorption spectrum of TFB computed using the vertical-Hessian thawed Gaussian approximation (``VH TGA''), compared with the experimental spectrum.}
    \label{fig:TFB}
\end{figure}

\bibliography{biblio52,additions_DoubleWell}